\documentclass[conference]{IEEEtran}
\usepackage{setspace}
\usepackage{clrscode}
\usepackage{amsmath}
\usepackage{amssymb}
\usepackage[dvips]{graphicx}%
\usepackage{epsfig}
\usepackage{hyperref}
\usepackage{bm}
\usepackage{bbm}
\usepackage{subfigure}
\usepackage{indentfirst}
\usepackage{float}
\usepackage{balance}
\usepackage{caption}
\usepackage{booktabs}

\newcommand{\figsize}{0.38}

\newtheorem{Def}{Definition}
\newtheorem{Prob}{Problem}

\begin{document}

\title{Age-Energy Tradeoff in Fading Channels with Packet-Based Transmissions}
    \author{
\IEEEauthorblockN{Haitao Huang\IEEEauthorrefmark{1}, Deli Qiao\IEEEauthorrefmark{1} and M. Cenk Gursoy\IEEEauthorrefmark{2}}
\IEEEauthorblockA{\IEEEauthorrefmark{1}\small{School of Communication and Electronic Engineering, East China Normal University, Shanghai, China 200241}}
\IEEEauthorblockA{\IEEEauthorrefmark{2}\small{Department of Electrical Engineering and Computer Science, Syracuse University, NY 13210}}
\small{Email: 51181214032@stu.ecnu.edu.cn, dlqiao@ce.ecnu.edu.cn, mcgursoy@syr.edu}}

\maketitle

\begin{abstract}
The optimal transmission strategy to minimize the weighted combination of age of information (AoI) and total energy consumption is studied in this paper. It is assumed that the status update information is obtained and transmitted at fixed rate over a Rayleigh fading channel in a packet-based wireless communication system. A maximum transmission round on each packet is enforced to guarantee certain reliability of the update packets. Given fixed average transmission power, the age-energy tradeoff can be formulated as a constrained Markov decision process (CMDP) problem considering the sensing power consumption as well. Employing the Lagrangian relaxation, the CMDP problem is transformed into a Markov decision process (MDP) problem. An algorithm is proposed to obtain the optimal power allocation policy. Through simulation results, it is shown that both age and energy efficiency can be improved by the proposed optimal policy compared with two benchmark schemes. Also, age can be effectively reduced at the expense of higher energy cost, and more emphasis on energy consumption leads to higher average age at the same energy efficiency. Overall, the tradeoff between average age and energy efficiency is identified.
\end{abstract}

\section{Introduction}
With emerging delay-sensitive applications, the timely update of the packet-based information is becoming increasingly more important. For instance, in status update systems such as in family care, security alert and environment monitoring scenarios, keeping system information fresh is extremely critical and essential \cite{IoT}. Age of information (AoI) has been proposed to measure the freshness of the status information in \cite{how often}. Specifically, AoI is defined as the time that has elapsed since the generation of last successfully received system information.

Recently, AoI has attracted much interest from the academia \cite{zero-wait}-\cite{mobile pushing}. For instance, general update policies such as zero-wait policy have been studied in \cite{zero-wait}, where an efficient optimal update mechanism has been designed. Under the assumption of using ``generate-at-will" model, by employing the queueing theory, how the packets should be managed in the buffer aided scheme has been addressed in \cite{queue}. The authors mainly concentrated on the system performance under M/M/1 and M/M/1/2 queuing systems with first-come-first-served (FCFS) policies. Poisson arrival processes with last-generated-first-served (LGFS) strategy have been considered in \cite{LGFS}, as an extension from the last-come-first-served (LCFS) system. The limited nature of the transmission power or the total energy in communication systems has necessitated the age minimization problems to be addressed under energy constraints.  For instance, considering energy harvesting devices and batteries, the authors have proposed an energy-aware adaptive status update policy for the information refresh mechanism in \cite{harvest1}. The authors  in \cite{harvest2} have considered an optimal sensing scheduling policy for energy harvesting sensing system and discussed the system performance with finite and infinite battery sizes separately. Also, the performance of the proposed policy was shown to match the theoretical bounds. The integer battery policies were generalized in \cite{harvest tradeoff}, and the threshold polices have been characterized. It is worth noting that the minimum average AoI is equivalent to the optimal threshold for the highest energy state.

Meanwhile, retransmission of the update packets should be considered in certain cases as well. A perfect feedback channel was used in \cite{update failure}, and a status updating policy called best-effort uniform updating with retransmission has been proposed. This update policy intends to equalize the delay between two successful updates.  Considering the hybrid ARQ (HARQ) protocol, the peak-age optimization problem of short packets has been explored in \cite{reliable}. Additionally, assuming that HARQ is combined with three kinds of actions at the source node, the average age minimization problem subject to a retransmission upper bound constraint has been solved in \cite{E's Journal}. Investigating how the upper bound on the transmission round alters the average age, the authors  in \cite{retrans} have shown that a higher upper bound would generally cause higher average age . The tradeoff between average age and normalized energy consumption was also demonstrated in \cite{retrans}, where both sensing energy and transmission energy have been taken into account. The age-energy tradeoff was also studied in \cite{mobile pushing}, where the tradeoff has been analyzed statistically by introducing a tradeoff parameter.

We note that the transmission power is fixed in most prior work. In this paper, we consider a joint optimization problem under an average power constraint. Different from previous work described above,  we consider the case in which the transmission power varies with the status of the packet. We first formulate a CMDP problem considering the weighted sum of age and total power consumption, and then use the Lagrangian method to transform the formulated CMDP problem into an MDP problem. We propose a new algorithm to derive the optimal policy. Through simulation results, we show the effectiveness of the proposed policy.

The rest of this paper is organized as follows, Section \uppercase\expandafter{\romannumeral2} discusses preliminaries corresponding to system model, AoI and energy efficiency. In Section \uppercase\expandafter{\romannumeral3}, the CMDP optimization problem is formulated. In Section \uppercase\expandafter{\romannumeral4}, the solution to the optimization problem is detailed. Numerical results are provided in Section \uppercase\expandafter{\romannumeral5}, while Section \uppercase\expandafter{\romannumeral6} concludes this paper.
\section{Preliminaries}

\subsection{System Model}

\begin{figure}
    \centering
    \includegraphics[width=\figsize\textwidth]{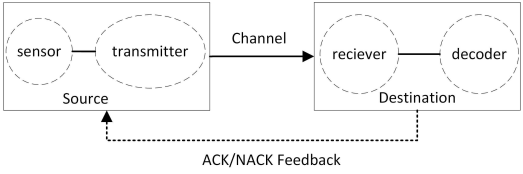}
    \caption{System model.}
    \label{fig:systemmdoel}
\end{figure}

We consider a wireless communication system with a feedback mechanism as shown in Fig. \ref{fig:systemmdoel}. In this model, the transmitter sends the packet to the receiver once the new message is generated by the sensor. Define $T$ as the duration of each transmitting time slot. Block fading channel model is assumed in this paper, i.e., the channel gain keeps constant in each time slot and  varies independently over different time slots. We further assume that feedback is transmitted over an error-free channel instantly. The number of transmission  rounds for one update packet is supposed to be $M$, and such retransmissions are important  to guarantee a certain level of reliability for the status updates over fading channels. For instance, if the packet is decoded without error, the source node will receive an ACK as feedback and generate a new packet containing the latest system information and time stamp. If decoding failure occurs, the destination will send a NACK to the source in order to request one more transmission of the same packet. Note that, the same packet could be transmitted no more than $M$ times, i.e., once the source node gets NACK $M$ times consecutively, the packet should be discarded and a new packet is generated. Without loss of generality, $T$ is set to be 1 in this paper.

\subsection{Age of Information (AoI)}

Age of information (AoI) reflects the freshness of the information on the status update system. It is defined as the time elapsed since the generation of the last successfully received message containing system information. Suppose that the system is now at time $t$, and the last successfully received packet has the time stamp $U(t)$, then the age can be written as follows:
\begin{align}
\Delta(t) &= t-U(t).
\end{align}

\begin{figure}
    \centering
    \includegraphics[width=\figsize\textwidth]{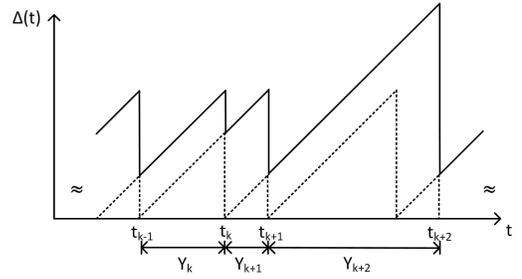}
    \caption{Evolution of age of information.}
   \label{fig:aoimodel}
\end{figure}

Fig. \ref{fig:aoimodel} illustrates how the AoI evolves in our system, where $t_k$ is time points when $k$-th packet is successfully transmitted. Denote $Y_k=t_k-t_{k-1}$ as the time interval between two consecutive packet generations. $Y_k$ can be seen to be actually equal to the transmission rounds for the $k$-th packet by noticing that time slot $T$ is normalized as $T=1$. When the packet is successfully received, the age is reduced. Otherwise, the age will keep increasing until the next ACK is obtained by the source node. The average age of information is given by
\begin{align}
\overline{\Delta}=\lim_{\tau\to\infty}\frac{1}{\tau}\int_{0}^{\tau}\Delta(t)dt.\label{eq:avagedef}
\end{align}

The calculation of the integration shown in (\ref{eq:avagedef}) can be explained as the area under the sawtooth curve in Fig. \ref{fig:aoimodel}, where the triangles in dashed line are isosceles triangles. Clearly, the area under the curve is the mixture of multiple trapezoids. Hence, the expression of average AoI can be rewritten as follows:
\begin{align}
\overline{\Delta}=\lim_{N\to\infty}\frac{\sum_{k=1}^{N}{(2Y_{k-1}+Y_{k})Y_{k}}}{2\sum_{k=1}^{N}Y_k}, \text{with} Y_0=0,\label{eq:avagedef1}
\end{align}
where $N$ is the total number of trapezoids. Given that $Y_k$ in (\ref{eq:avagedef1}) is composed of time slots for transmission, AoI can be measured by the packet transmission rounds.

\subsection{Energy Efficiency}
In the sensor-equipped communication system, the energy cost can be divided into two parts: sensing usage and transmission consumption. The energy for sensing is used to obtain the newest status information, and the energy for transmitting the packet via fading channel is needed in each round of transmission. For the clarity of description, several definitions are given in the following:
\begin{Def}
Define $C_{max}$ as the constraint on the average transmission power that the wireless system could use.
\end{Def}
\begin{Def}
Define $P_{tk}$ as the total transmission power that the source node uses to deliver the $k$-th information packet regardless of having success or not, from which the sensing power is excluded.
\end{Def}
\begin{Def}
Define $\alpha$ as the ratio between $C_{max}$ and sensing power $P_s$, i.e., $P_s$=$\alpha$$\cdot$$C_{max}$.
\end{Def}

Energy efficiency represents the fraction of power in total consumption for sending packets considering sensing power cost. Based on the above definitions, the energy efficiency can be calculated as
\begin{align}
\psi = \frac{\sum_{k=1}^{n}P_{tk}}{\sum_{k=1}^{n}P_{tk}+n \cdot P_s},\label{eq:efficiency}
\end{align}
where $n$ is the number of sensing times in all transmission rounds, i.e., the total number of update packets generated. From equation (\ref{eq:efficiency}), we note that generating a new packet affects the energy efficiency apparently, and more frequent generation and higher sensing power result in lower energy efficiency. By measuring energy efficiency, one can determine whether the system performs efficiently. This also enables us  to compare the effectiveness of different transmission strategies as discussed later.

\begin{figure}
    \centering
    \includegraphics[width=\figsize\textwidth]{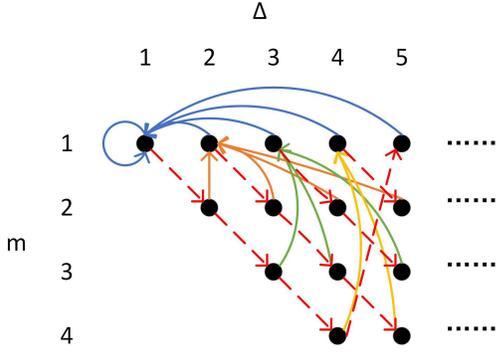}
    \caption{Markov chain model with retransmit actions.}
   \label{fig:retansmdoel}
\end{figure}

\begin{figure}
    \centering
    \includegraphics[width=\figsize\textwidth]{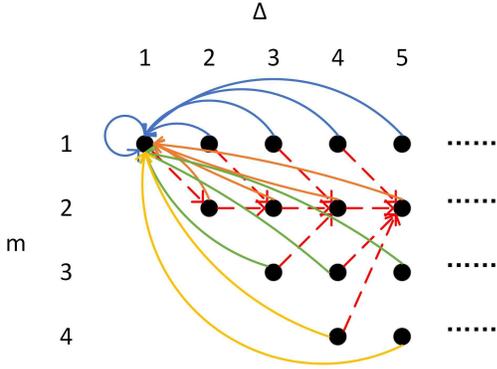}
    \caption{Markov chain model with generate and transmit actions.}
   \label{fig:generatemdoel}
\end{figure}
\section{Formulation of CMDP Problem}

In this section, we first discuss the definitions of four tuples in the CMDP problem: states, actions, reward and transition probabilities. Subsequently,  we formulate the CMDP problem to optimize the average age and weighted power consumption.

\subsection{States and Actions}
In this paper, a specific state can be determined by a certain age-transmission pair ($\Delta,m$). The state space $\mathcal{S}$ is in general countably infinite, but points with $m>\Delta$ should be excluded from the state space since age cannot be smaller than the number of transmission rounds. Also, according to the retransmission scheme described in the system model subsection, the value of $m$ should be less than its upper-bound $M$. So the state space  can be expressed as
\begin{align}
\mathcal{S}=\{(\Delta,m):m\leq\min\{\Delta,M\},\Delta\in\mathcal{D},m\in\mathcal{M}\}, \nonumber
\end{align}
where $\mathcal{D}$ = $\{1,2,3...\}$ is the set of age values that the system can reach, $\mathcal{M}$ = $\{1,2,..,M\}$ is set of transmission rounds for a packet.

\begin{figure}
    \centering
    \includegraphics[width=\figsize\textwidth]{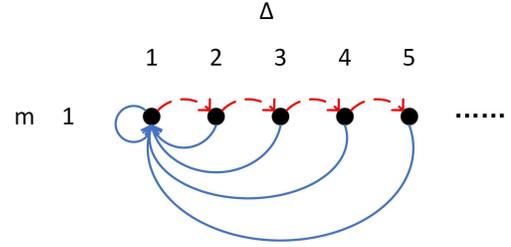}
    \caption{Markov chain model of keeping generation action.}
   \label{fig:allsensemodel}
\end{figure}

The action set is given by $\mathcal{A} = \{r,g\}$: (\romannumeral1) retransmit the packet failed previously with power consumption $P_t$ ($a=r$); (\romannumeral2) generate and transmit a new packet with power consumption $P_s+P_t$ ($a=g$), and $P_t$ can take various values from a set that will be explained later for both (\romannumeral1) and (\romannumeral2). Different kinds of policies result in different Markov chain models as shown in Fig. \ref{fig:retansmdoel} and Fig. \ref{fig:generatemdoel} correspondingly. The solid lines and dashed lines respectively represent successful and failed transitions. Let us consider Fig.\ref{fig:retansmdoel} with retransmit action, and suppose the update packet is not correctly received with $a(\Delta,m)$ = $r$. Then, the system will step into state $(\Delta+1,m+1)$, $m<M$. If the transmission round  $m$ reaches the upper-bound, a new packet should be generated, which causes the system to go to state $(\Delta+1,1)$. On the other hand, considering  Fig. \ref{fig:generatemdoel} with generate and transmit action, if the system generates a new packet, the unsuccessful reception will lead to a transition from state $(\Delta,m)$ to state $(\Delta+1,2)$, since the generated packet has been transmitted once. Meanwhile, if the transmission is successful, the state will change to $(m,1)$ and $(1,1)$ with the actions $r$ and $g$, respectively. Moreover, if the source node chooses to generate a new packet in each time slot, the equivalent Markov chain model is shown in Fig. \ref{fig:allsensemodel}. Since the packet to be sent in every time slot is brand new, the transmission round will remain at one, and transmission failure only  increases the AoI. Additionally, the age is reduced to one from each state in this mechanism when a successful transmission occurs.

\subsection{Transition Probability}
We assume that the channel distribution information is known at the source node. Suppose that the channel is quantized into $K$ levels. Let us denote the channel gain set as $\mathcal{G}$ =$\{z_0,z_1,..,z_{K}\}$ as the channel gain set and assume that the elements in $\mathcal{G}$ are organized in increasing order with $z_{0}=0$ and $z_{K}=\infty$. Suppose that the channel is now in state $j$ with the channel gain $z$ $\in$ [$z_j, z_{j+1}$). Once the channel state $j$ transits into state $i$, the block  fading channel assumption leads to
\begin{align}
\Pr\{z_i|z_j\} = \int_{z_i}^{z_{i+1}} p_z(z)dz = \beta_i,
\end{align}
where $p_z(z)$ is the probability density function of the channel gain, and $\beta_i$ is the probability of channel state $i$. According to Shannon's formula \cite{fundamental}, the power of $k$-th level can be specified by the following formula assuming unit noise power:
\begin{align}
P_k=\frac{2^R-1}{z_{k}}, k=1,2,..K
\end{align}
where $R$ represents the data rate or packet size equivalently. Depending on the system status, different states might result in various transmission power levels. Hence, the transmission power $P_t$ belongs to the power set $\{P_1,P_2,...,P_K\}$. The probabilities of unsuccessfully receiving a packet given transmission power $P_k$ are expressed as
\begin{align}
\epsilon_k=\sum_{i=1}^{k} \beta_i.
\end{align}

To sum up, the state transition probability for the formulated Markov chain can be listed as follows:

\noindent{$Retransmission$ $action:$}
\begin{align}
{\rm Pr}(\Delta+1,m+1|\Delta,m,r)=&\epsilon_k, m < M\\
{\rm Pr}(\Delta+1,1|\Delta,m,r)=&\epsilon_k, m = M\\
{\rm Pr}(m,1|\Delta,m,r)=&1-\epsilon_k,
\end{align}
$Generate$ $and$ $transmit$ $action:$
\begin{align}
{\rm Pr}(\Delta+1,2|\Delta,m,g)=&\epsilon_k,\\
{\rm Pr}(1,1|\Delta,m,g)=&1-\epsilon_k,
\end{align}
$Both$ $actions$ $:$
\begin{align}
{\rm Pr}(\Delta^{\prime},m^{\prime}|\Delta,m,A)=&0, {\rm otherwise.}
\end{align}

\subsection{Reward and Problem Formulation}
Define the decision $\mu_t(s)$ as a mapping that connects the system state $s=(\Delta_t,m_t)$ to the action to be taken at  time $t$. Therefore, $s_t^{\mu}=(\Delta_t^{\mu},m_t^{\mu})$ denotes the sequences induced by policy $\mathbf{\mu}$ at each $(\Delta_t,m_t)$. If the set of decisions $\mathbf{\mu}=\{\mu_1,\mu_2,...\}$ is not dependent on the time slot $t$, the policy can be called as  a stationary policy. For a stationary policy $\mathbf{\mu}$, the long term average AoI $\overline{\Delta}$ and power consumption $\overline{P}$ have the following forms:
\begin{align}
\overline{\Delta}(s,\mathbf{\mu})&=\sum_{t=1}^{\infty}\gamma^{t-1}\mathbb{E}\{\Delta_t^\mu+\frac{1}{2}\},\label{eq:avaoi}\\
\overline{P}(s,\mathbf{\mu})&=\sum_{t=1}^{\infty}\gamma^{t-1}\mathbb{E}\{p_t^{\mu}\},
\end{align}
where $\gamma\in(0,1)$ is a discount factor and the expectations are taken over the policy $\bm{\mu}$. Note that as $\gamma\to1$, the infinite sum of the discounted rewards and costs converge to their corresponding expected average rewards and costs \cite{Qiao17}. The long term average transmission power consumption $\overline{C}$ or $cost$ function is
\begin{align}
\overline{C}(s,\mathbf{\mu})=\sum_{t=1}^{\infty}\gamma^{t-1}\mathbb{E}\{p_t^{\mu}-P_s \cdot \mathbbm{1}[a=g]\},\label{eq:avcost}
\end{align}
where $p^{\mu}_t$ is the power used in time slot $t$ with policy derived form $\mu$ and the $p^{\mu}_t$ belongs to the power set \{$P_1$,$P_2$,...,$P_K$\} if the source node takes a retransmit action or \{$P_s+P_1$,$P_s+P_2$,...,$P_s+P_K$\} if a new packet is generated and transmitted. Symbol $\mathbbm{1}[\cdot]$ in (\ref{eq:avcost}) is an indicator function, whose value equals to one when action $g$ is taken. In order to investigate the tradeoff between age and total power consumptions, the $reward$ function of the CMDP problem is defined as:
\begin{align}
R(s,\mathbf{\mu})=\overline{\Delta}(s,\mu)+\omega\overline{P}(s,\mu),
\end{align}
where $\omega>0$ is a weight  factor. The reward function has combined the age and total power consumption under policy $\mu$ of a certain state, which satisfies our joint optimization aim, so the formulated CMDP problem can be expressed as:
\begin{Prob}\label{prob1}
\begin{align}
\min_{\mu} \,\,&R(s,\mathbf{\mu})\hspace{1cm}
\text{s.t.} \,\,\overline{C}(s,\mu) \leq C_{max}\nonumber
\end{align}
\end{Prob}
where $C_{max}$ is maximum average transmission power.

\section{Optimal Policy}
In this section, Lagrangian relaxation is used to solve the problem given in the former section, and an algorithm is proposed.

Note that the state space is countably infinite because of the arbitrary values of AoI, but in practice a large finite space is employed. This simplification is due to the infinite state space being approximated by a finite state space having a large upper bound $\Delta_{max}$. Whenever the AoI increases beyond $\Delta_{max}$, we set it to be one. In this way, a finite countable state Markov decision progress (MDP) approximation problem could be solved \cite{E21}. The optimal policy for the simplified MDP problem still converges to the policy for the original CMDP problem. Lagrangian relaxation for   Problem {\ref{prob1}} is carried out with non-negative multiplier $\eta$, and $Lagrangian$ $reward$ for the problem is defined as
\begin{align}
r(s,\mu_,\eta)=\Delta(s,\mu)+\frac{1}{2}+\omega p(s,\mu)+\eta c(s,\mu),\label{eq:newreward}
\end{align}
where $ p(s,\mu)$ and $c(s,\mu)$ denote the total power and transmit power consumption in state $s$ with policy $\mu$, respectively. Then, the Bellman optimality equation for (\ref{eq:newreward}) is given by
\begin{align}
V_\eta(s)=\min_{a\in A} \bigg\{r(s,\mu,\eta)+\gamma \sum_{s'\in S}{\rm Pr} \{s'|s,a\}V_\eta(s')\bigg\}.\label{eq:Bellman}
\end{align}
Note that Problem {\ref{prob1}} can be transformed to the following Problem {\ref{prob2}} with the given $\eta$ and $\gamma$:
\begin{Prob}\label{prob2}
\begin{align}
\min_{a\in A}\,\,&V_{\eta}(s,a)\hspace{1cm}
\text{s.t.} \,\,\overline{C}(s,\mu) \leq C_{max} \nonumber
\end{align}
\end{Prob}
Consequently, the action set for the minimized $V_{\eta}(s,a)$ is
\begin{align}
\mu_{\eta}^* \triangleq \arg\min_{a\in A}V_\eta(s,a).
\end{align}

As a widely used method, Value Iteration Algorithm (VIA) is adequate to solve this problem. With the initialization of arbitrary state $s$, $V_{\eta}^0(s)$ and Lagrangian multiplier, $V_{\eta}^{n+1}(s)$ will be updated in each iteration until convergence. Hence, given $\eta$, the steady state probability $\boldsymbol{\pi}_{\eta}$ of the Markov chain can be calculated, and we can rewrite ({\ref{eq:avaoi}})-({\ref{eq:avcost}}) as
\begin{align}
\overline{\Delta}_{\eta}&=\sum_{s}\Delta(s)\boldsymbol{\pi}_{\eta}(s),\\
\overline{P}_{\eta}&=\sum_{s}p(s)\boldsymbol{\pi}_{\eta}(s),\\
\overline{C}_{\eta}&=\sum_{s}c(s)\boldsymbol{\pi}_{\eta}(s),
\end{align}
where $c(s)$ is the transmission power used in state $s$ and $p(s)$ is the total power consumed in states. As indicated in \cite{E's Journal}, \cite{Qiao16} and \cite{Qiao17}, the structure of the optimal policy should be the mixture of two stationary polices. Obviously, $\overline{C}_\eta$ is monotonically decreasing when $\eta$ is increasing. Thus, bisection method can be used to find the optimal $\eta$. The optimal policy $\mu_{\eta}^*(s)$ always chooses policy $\mu_{\eta^-}(s)$ with probability $\xi$ and $\mu_{\eta^+}(s)$ with probability $1-\xi$. The mixture coefficient $\xi$ can be obtained by solving $\xi\overline{C}_{\eta^-}+(1-\xi)\overline{C}_{\eta^+}=\overline{C}$. The detailed description of Algorithm 1 is provided below.
\begin{table}[htbp]
\begin{center}
 \begin{tabular}{lcl}
 \toprule
 \textbf{Algorithm 1: Optimal policy }   \\
 \midrule
  \textbf{Input:} $C_{max},K,M,\Delta_{max},R,\gamma,P_k,P_s$;\\
  \textbf{Specify:} $\zeta_{\eta},\zeta_V$;\\
  \textbf{Initialize:} $V_0^{\mathcal{S}}\leftarrow\mathbf{0},n=0,\eta^-=0,\eta^+$;\\
  \textbf{While} $|\eta^{+}-\eta^{-}|>\zeta_{\eta}$;\\
  \textbf{do} $\eta=(\eta^-+\eta^+)/2$;\\
  \hspace{0.2cm}\textbf{While} $\max_s|V_{n+1}-V_{n}|>\zeta_V$\\
  \hspace{0.2cm}\textbf{do}\\
  \hspace{0.4cm}$V_{n+1}(s)=\mathop{min}\limits_{a\in A}\{(1-\gamma)(\Delta+0.5+\omega P+\eta P_k)$\\
  \hspace{2.4cm}$+\gamma\mathbb{E}[V_n(s')]\}$;\\
  \hspace{0.4cm}$n=n+1$;\\
  \hspace{0.2cm}\textbf{End}\\
  \hspace{0.2cm}Derive the policy $\mu_{\eta}(s)=arg \mathop{min}\limits_{a\in A} V_\eta(s)$;\\
  \hspace{0.2cm}Calculate the corresponding steady state distribution \\
  \hspace{0.2cm}$\boldsymbol{\pi}_{\eta}(s)$ and average cost $\overline{C}_\eta$;\\
  \hspace{0.2cm}\textbf{If} $\overline{C_\eta}>C_{max}$\\
  \hspace{0.2cm}$\eta^-=\eta$;\\
  \hspace{0.2cm}\textbf{else}\\
  \hspace{0.2cm}$\eta^+=\eta$;\\
  \hspace{0.2cm}\textbf{End}\\
  \textbf{End}\\
Compute $\overline{\Delta}_{\eta^+},\overline{\Delta}_{\eta^-},\overline{C}_{\eta^+},\overline{C}_{\eta^-}$ and $\xi$;\\
Let the optimal average AoI be:\\
$\overline{\Delta}=\xi\overline{\Delta}_{\eta^-}+(1-\xi)\overline{\Delta}_{\eta^+}$.\\
 \bottomrule
 \end{tabular}
 \end{center}
 \end{table}

\section{Numerical Results}
In this paper, we assume that the channel state probabilities are $\beta_1=\beta_2=...=\beta_K=\frac{1}{K}$, $K=128$ and wireless channels are Rayleigh fading channels with unit mean. What is more, $M=4$, $\Delta_{max}=100$ and $R=1$bps/Hz.
\begin{figure}
    \centering
    \includegraphics[width=\figsize\textwidth]{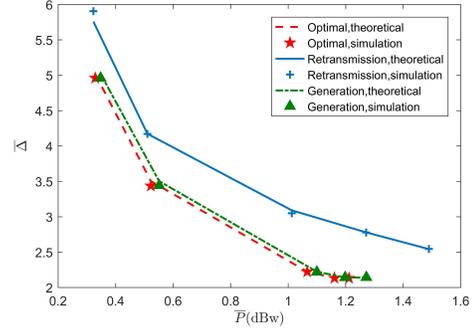}
    \caption{Age-energy tradeoff with different policies.}
   \label{fig:multipolicy}
\end{figure}

In Fig. \ref{fig:multipolicy}, we plot the average age as the average transmit power varies from $-5$dBw to $3$dBw. We assume $\omega=1$ for the proposed policies. In the figure, ``Optimal'' is the one proposed in this paper, ``Retransmission'' refers to the policy that each packet is retransmitted until success or reaching the maximum transmission round $M$, and ``Generation'' is the policy that a new packet is generated in each time slot. First, we can see from the figure that the simulation results match with the theoretical values calculated from the analysis. Also, the proposed policy achieves the same average age with the lowest total power consumption in all cases, which indicates higher energy efficiency.

\begin{figure}
    \centering
    \includegraphics[width=\figsize\textwidth]{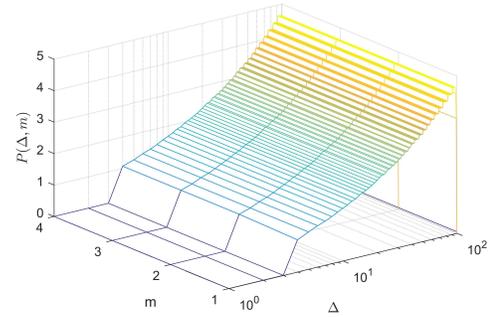}
    \caption{Transmission power policy.}
   \label{fig:power}
\end{figure}

In Fig. \ref{fig:power}, we plot the power allocation policy for different states when $\omega=1$ and $C_{max}=-3$dBw. It is shown in the figure that the source node keeps silent when both age and the number of transmission round are relatively small, then with the increase of AoI, the source node uses higher power to achieve successful transmissions.

\begin{figure}
    \centering
    \includegraphics[width=\figsize\textwidth]{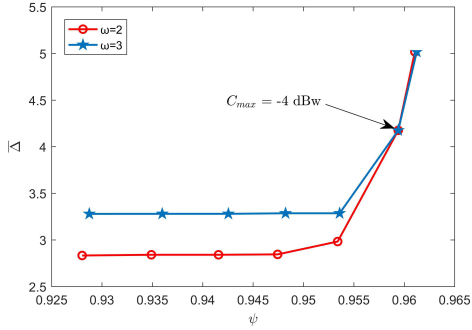}
    \caption{Average AoI and energy efficiency tradeoff under various values of $\omega$.}
   \label{fig:efficiency}
\end{figure}

In Fig. \ref{fig:efficiency}, we plot the average age in terms of energy efficiency as average power constraint decreases form $0$dBw to $-5$dBw. In this figure, the average age for $\omega=2$ is always smaller when compared with $\omega=3$. This is basically because when $\omega$ is larger, the system views power minimization more important. Similarly, the system can achieve higher energy efficiency under the same power constraint if tradeoff parameter $\omega$ is larger.

\begin{figure}
    \centering
    \includegraphics[width=\figsize\textwidth]{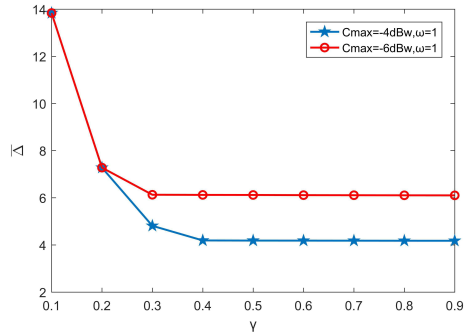}
    \caption{Average AoI as a function of discount factor.}
   \label{fig:multidis}
\end{figure}

In Fig. \ref{fig:multidis}, we plot the average age as a function of the discount factor $\gamma$. The discount factor implies how much the system cares about future reward or long term benefit. The larger value of $\gamma$ means the greater degree of future reward being considered in optimization. Obviously, when $\gamma$ is large, the system achieves the average age.

\begin{figure}
    \centering
    \includegraphics[width=\figsize\textwidth]{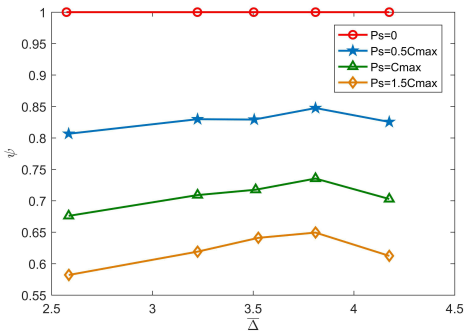}
    \caption{Impact of ratio factor in age-energy tradeoff.}
   \label{fig:multialpha}
\end{figure}

In Fig. \ref{fig:multialpha}, we investigate the tradeoff between average age and energy efficiency with different $\alpha$. Given $\omega=1$, as $C_{max}$ decreases from $-1$dBw to $-4$dBw, we can find that average age keeps increasing, while the energy efficiency improves at first and then gets worse. This is because with the increase in age, less frequent generation of update packets can be expected, which improves energy efficiency. But when average transmission power is below a threshold, more packets should be generated to reduce the age effectively, and hence energy efficiency is reduced.

\section{Conclusion}\label{NR}

In this paper, we have investigated the age-energy tradeoff in fading channels with packet-based transmissions. We have assumed that the packets are transmitted at fixed rate and the transmission power levels can be adjusted for different system states. Afterwards, a CMDP problem has been formulated to optimize the weighted sum of average age and total power consumption. We have adopted the Lagrangian method to relax the CMDP problem to an MDP problem and solved the specific optimization problem using Bellman optimality equations. In particular, we  have proposed an algorithm to solve the problem and obtain the best power allocation policy. According to the simulation results, the policy obtained from the proposed algorithm could reduce average age and power consumption simultaneously. Also, more emphasis on the energy consumption leads to large average AoI at the same energy efficiency.


\begin{thebibliography}{1}

\bibitem{IoT} M. Elmagid, N. Pappas and H. Dhillon, ``On the role of age of information in the internet of things,'' \emph{IEEE Commun. Magazine}, pp. 72 - 77, Dec. 2019.

\bibitem{how often} S. Kaul, R. Yates and M. Gruteser, ``Real-time status: How often should one updates?'' IEEE INFOCOM 2012.

\bibitem{zero-wait} Y. Sun, E. Biyikoglu and R. Yates, ``Update or wait: How to keep your data fresh,'' \emph{IEEE Trans. Inform. Theory}, vol. 63, no. 11, pp. 7492 - 7508, Nov. 2017.

\bibitem{queue} M. Costa, M. Codreanu and A. Ephremides, ``On the age of information in status update systems with packet management,'' \emph{IEEE Trans. Inform. Theory}, vol. 63, no. 4, pp. 1897 - 1910, Apr. 2016.

\bibitem{LGFS} A. Bedewy, Y. Sun and N. Shroff, ``Minimizing the age of information through queues,'' \emph{IEEE Trans. Inform. Theory}, vol. 65, no. 8, pp. 5215 - 5232, Aug. 2019.

\bibitem{harvest1} H. Wu, J. Yang, and J. Wu, ``Optimal status update for age of information minimization with an energy harvest source,'' \emph{IEEE Trans. Green. Commun. Netw.}, vol. 2, no. 1, pp. 193 - 204, Mar. 2018.

\bibitem{harvest2} J. Yang, X. Wu and J. Yang, ``Optimal online sensing scheduling for energy harvesting sensor with infinite and finite batteries,'' \emph{IEEE J. Sel. Areas Commun.}, vol. 34, no. 5, pp. 1578 - 1589, May. 2016.

\bibitem{harvest tradeoff} B. Bacinoglu, Y. Sun, E. Biyikoglu and V.Mutlu, ``Achieving the age-energy tradeoff with a finite-battery energy harvesting source,'' IEEE ISIT 2018.

\bibitem{update failure} S. Feng and J. Yang, ``Minimizing age of information for an energy harvesting source with update failures,'' IEEE ISIT 2018.

\bibitem{reliable} R. Devassy, G. Durisi, G. Ferrante, O. Simeone and E. Uysal ``Reliable transmission of short packets through queues and noisy channels under latency and peak-age violation guarantees,'' \emph{IEEE J. Sel. Areas Commun.}, vol. 37, no. 4, pp. 721 - 734, Apr. 2019.

\bibitem{E's Journal} E. Ceran, D. Gunduz and A. Gyorgy, ``Average age of information with hybrid ARQ under a resource constraint,'' \emph{IEEE Trans. Wireless Commun.}, vol. 18, no. 3, pp. 1900 - 1913, Mar. 2019.

\bibitem{retrans} J. Gong, X. Chen and X. Ma, ``Energy-age tradeoff in status update communication systems with retransmission,'' IEEE GLOBECOM 2018.

\bibitem{mobile pushing} S. Nath, J. Wu and J. Yang, ``Optimizing age-of-information and energy efficiency tradeoff for mobile pushing notifications,'' IEEE SPAWC 2018.

\bibitem{fundamental} D. Tse and P. Viswanath, ``Fundamentals of Wireless Communication,'' Cambridge Univ. Press, 2005.

\bibitem{E21} L. I. Sennott, ``Constrained average cost Markov decision chains,,'' \emph{Probab. Eng. Inf. Sci.}, vol. 7, no. 1, pp. 69 - 83, Jan. 1993.

\bibitem{Qiao16} E. Altman, `` Constrained Markov Decision Processes,'' Chapman and Hall\&CRC, 1998.

\bibitem{Qiao17} M. H. Ngo and V. Krishnamurthy, ``Monotonicity of constrained optimal transmission policies in correlated fading channels with ARQ,'' \emph{IEEE Trans. on Signal Processing.}, vol. 58, no. 1, pp. 438 - 451, Jan. 2010.

\end{thebibliography}
\end{document}